\documentstyle[prl,aps,amsfonts,amssymb,psfig,multicol]{revtex}

\begin{document}
\draft
\preprint{{\bf ETH-TH/98-??}}

\title {Exact free energy distribution function of a randomly
forced directed polymer}

\author{D. A. Gorokhov and G. Blatter}

\address{Theoretische Physik, ETH-H\"onggerberg, CH-8093 Z\"urich,
Switzerland\\
e-mail: gorokhov@itp.phys.ethz.ch}

%\date{\today}
\maketitle
\begin{abstract}
{ We study the elastic ($1+1$)-dimensional string subject to a random
gaussian potential on scales smaller than the correlation radius of
the disorder potential (Larkin problem). We present an exact
calculation of the probability function ${\cal P}\left
[F\left(u,L\right)\right]$ for the free energy $F$ of a string
starting at $(0,0)$ and ending at $(u,L)$. The function ${\cal P}(F)$
is strongly asymmetric, with the left tail decaying exponentially
($\ln {\cal P}(F\rightarrow-\infty)\propto F$) and the right tail
vanishing as $\ln {\cal P}(F\rightarrow +\infty)\propto -F^{3}$. Our
analysis defines a strategy for future attacks on this class of
problems. }

\end{abstract}


\begin{multicols}{2}  
\narrowtext

Over the recent years, the interest in static and dynamic aspects of
random manifold problems \cite{Halpin} has increased significantly,
with numerous applications in the field of random
magnets\cite{Forgacs}, vortex matter in type II superconductors
\cite{Blatter}, dislocations in metals \cite{Ioffe}, charge-density
waves in solids \cite{Gruner}, and many more. Also, the physics of
random elastic manifolds is related to other fascinating topics, e.g.,
Burgers turbulence \cite{Mezard}, stochastic growth \cite{Krug}, or
systems of interacting bosons \cite{Lieb}, all of which have attracted
considerable interest recently. In this context, the Larkin model
\cite{Larkin} has become a generic testing-field for various ideas,
similar to the harmonic oscillator in quantum mechanics --- both
model systems are simple and exactly solvable, while at the same time
being very instructive.  Quite astonishingly, the study of the Larkin
model has not been carried through in all its facets.  In particular,
the distribution function ${\cal P}(F)$ of the free energy $F$, an
important element in the characterization of the dirty elastic
manifold, is not known to date. The present work eliminates this
deficiency with an exact calculation of ${\cal P}(F)$.

A $d$-dimensional dirty elastic manifold (elasticity $\epsilon$) with
$n$ transverse degrees of freedom ${\bf u}({\bf z})$ (${\bf u}\in
{\mathbb{R}}^{n},~{\bf z}\in {\mathbb{R}}^{d}$) is described by the
Hamiltonian
\begin{equation}
{\cal H}[{\bf u}({\bf z})]= \int d^d z \bigg[
\frac{\epsilon}{2}\bigg(\frac{\partial{\bf u}}{\partial{\bf
z}}\bigg)^2 + U({\bf u({\bf z})},{\bf z}) \bigg].
\label{Hamiltonian}
\end{equation}
The disorder potential $U$ is a random gaussian variable with a
correlator $K(|{\bf u}|)$ decaying on the scale $R_{\scriptscriptstyle
K}$,
\begin{equation}
\left \langle U({\bf u},{\bf z}) U({\bf u}^{\prime},{\bf z}^{\prime})
\right \rangle = K(|{\bf u}-{\bf u}^{\prime}|) \delta({\bf z}-{\bf
z}^{\prime}).
\label{correlr}
\end{equation}
The partition function $Z({\bf u},L)$ of a string ($d = 1$) beginning
and ending at the points $({\bf 0},0)$ and $({\bf u}, L)$ can be
written as a path integral
\begin{eqnarray}
&&Z ({\bf u}, L) \!\! = \!\!\int\limits_{({\bf 0},0)}^{({\bf u},L)}  
{\cal D}[{\bf u}^{\prime}({z}^{\prime})]
\, \exp\bigg\{\!\!-\frac{1}{T} \! \int_0^L \!\!\! dz^{\prime}
\bigg[ \frac{\epsilon}{2}\bigg(\frac{\partial{\bf u}^{\prime}}
{\partial{z^{\prime}}}\bigg)^2 \label{partitionf}\\ 
&& \hspace{5.8 cm} + U({\bf u}^{\prime}(z^{\prime}),{z}^{\prime}) 
\bigg] \bigg\},\nonumber
\end{eqnarray}
from which we obtain the free energy $F=- T\ln Z({\bf u},L)$.  Due to
the random nature of the potential $U$, the free energy $F$ is a
stochastic variable and we are interested in its distribution function
${\cal P}(F)$, see Fig.\ 1.

The imaginary time Schr\"odinger equation $T{\partial_z Z}=
(T^2/2\epsilon) \partial_{\bf u}^2 Z-UZ$, $z\in [0,L]$, satisfied by
the partition function $Z$, directly maps to the Kardar-Parisi-Zhang
equation\cite{KPZ} $\partial_z F = (T/2\epsilon) \Delta F - (\nabla
F)^2/2\epsilon + U$ describing the stochastic growth of the `surface'
$F({\bf u},z)$, and differentiating the KPZ-equation with respect to
${\bf u}$ and going over to the gradient field ${\bf v} = \nabla F$
leads to the Burgers equation \cite{Mezard}.
\begin{figure} [bt]
\centerline{\psfig{file=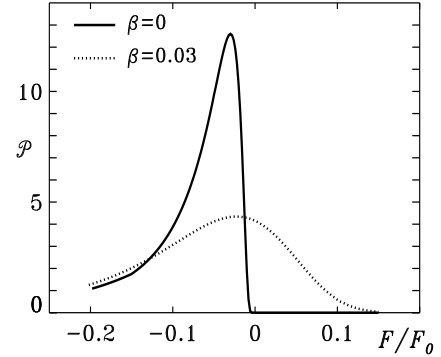,width=6cm,height=5cm}}
\narrowtext\vspace{2mm}
\caption{The function ${\cal P}(F)$ for the Larkin model (we have
subtracted the trivial thermal component $F_{\rm th}$ from $F$; the
parameters are $\beta = \epsilon^2 u^2/\alpha L^3$ and $F_0 = \alpha
L^2/ \epsilon$, with $\alpha =-K^{\prime\prime}(0)$ the disorder
parameter in the force correlator): the general case with $\beta \ne 0
\leftrightarrow u \ne 0$ (dotted line) is characterized by the tails
${\cal P}(F \rightarrow -\infty) \sim \exp(-\pi^2 |F|/F_0)$ and ${\cal
P}(F \rightarrow \infty)$ $\sim \exp(-16 F^3/27 \beta^2 F_0^3)$.  For
$\beta = 0 \leftrightarrow u=0$ (solid line) the distribution
function vanishes identically for $F>0$ with ${\cal P}(F \rightarrow
0^-) \sim \exp(-F_0/16 |F|)$. The distribution function for the
random polymer problem on short scales $u<R_{\scriptscriptstyle K}$, 
see Eqs. (1) and (2), is independent of $u$ and is given by the 
result for the Larkin model with $\beta = 0$.}
\end{figure}
In studying random elastic manifolds, we are interested in the
statistical behavior of the displacement field ${\bf u}$ and the free
energy $F$ as functions of the internal coordinates ${\bf
z}$. Important results for the directed polymer problem concern the
disorder induced line wandering $\langle{\bf u}^2\rangle \propto
L^{2\zeta}$ and the free energy average $\langle F\rangle$ and its
fluctuations. A more ambitious goal is the calculation of the
probability distributions ${\cal P}(u,L)$ and ${\cal P}(F,L)$.  Only
few {\it rigorous} results are known, among them the behavior of
$\langle{\bf u}^2\rangle$ and $\langle F\rangle$ of the Larkin
problem \cite{Parisi}, the solution of the $(1+1)$-directed polymer
problem by Huse, Henley, and Fisher\cite{HHF}, where $\zeta = 2/3$ is
an exact result, and the replica variational calculation for the
$(d+n)$-random manifold by Mezard and Parisi\cite{MP}, who find $\zeta
= (4-d)/(4+n)$ for $d>2$ and $\zeta = (2-d)/2$ for $d<2$, an exact
result in the limit $n\rightarrow\infty$.  The free energy probability
distribution function ${\cal P}(F)$ has been studied intensively (see
Refs.\ \cite{Kardar,Zhang,Kolomeisky,Maritan,Halpin1}) but few solid
results are available: the left tail $\ln {\cal P}(F\rightarrow
-\infty)\propto -(|F|^3/L)^{1/2}$ for the $(1+1)$-problem (from which
the free energy fluctuation exponent $\chi = 1/3$ follows trivially)
has been found by Kardar \cite{Kardar} and some (partly contradicting)
results for the tails of the velocity-difference correlation function
of the Burgers equation have been reported recently \cite{Polyakov}.

In this situation it is appropriate to look out for those problems
which can teach us some definitive non-trivial results.  Here, we
investigate the exactly solvable model of a random directed polymer on
scales smaller than the correlation radius $R_{\scriptscriptstyle K}$
of the random potential. The studies on this model go back to the
pioneering work of Larkin \cite{Larkin}, who, exploiting the smallness
of the displacement field ${\bf u}$, expanded the potential in the
Taylor series
\begin{equation}
U({\bf u},z)\approx U(0,z)-{\bf f}(z)\cdot {\bf u}.
\label{LM}
\end{equation}
This model is linear in ${\bf u}$ and thus allows for an exact
analytical solution. Though simple, it produces a number of
non-trivial results, of which the distribution function ${\cal P} (F)$
is still unknown.  Below, we solve the problem through a mapping on a
system of interacting bosons.

We start from the partition function (\ref{partitionf}) for the
$(1+1)$-dimensional string in a random force field $f(z)$ \cite{DR},
with $f(z)$ a random gaussian variable with correlator $\langle f(z)
f(z^\prime)\rangle =\alpha\delta (z-z^\prime)$, $\alpha = -K^{\prime
\prime}(0)$, see Eqs.\ (\ref{correlr}) and (\ref{LM}) [the 
generalization to the ($1+n$)-problem is trivial].  
Replicating the partition function $n$ times we obtain the 
Green function $G(u_1,\dots,u_n;L)\equiv \langle Z(u_1,L)\dots
Z(u_n,L) \rangle$ satisfying the imaginary time Schr\"odinger equation
($z\in [0,L]$)
\begin{equation}
T\frac{\partial G}{\partial z}=
\frac{T^{2}}{2\epsilon}\sum\limits_{i=1}^{n}
\frac{\partial^{2}G}{\partial u_{i}^{2}}
+\frac{\alpha}{2T}\sum\limits_{i,j=1}^{n}u_{i}u_{j}\thinspace G,
\label{green}
\end{equation}
with the initial condition $G (u_1,\dots,u_n;z\rightarrow +0)= \delta
(u_1)$ $\delta (u_2)\dots\delta (u_n)$. The boundary conditions at
infinity read $G(u_1,\dots,|u_{i}|\rightarrow\infty,\dots,u_n; z)
\rightarrow 0$. Equation (\ref{green}) has the form $T{\partial_{z}
G}=-{\hat H}G$, with ${\hat H}$ the Hamiltonian describing the
imaginary time evolution of a system of interacting bosons, with the
temperature $T$ playing the role of Planck's constant and the internal
coordinate $z$ of the string mapping on to the imaginary time
coordinate of the boson system. In the new variables $t = 
\sqrt{\alpha/2\epsilon T}\, z$ and $x_i=(\alpha\epsilon/2 T^3)^{1/4}\,
u_i$, Eq.\ (\ref{green}) takes the simple form $\partial_t G = (1/2) 
\sum_i \partial^2_i G + \sum_{i,j} x_i x_j \thinspace
G$. The specific form of the Hamiltonian motivates an Ansatz $G = g(t)
\exp[(1/2) \sum_{i,j} a_{ij}(t)x_i x_j]$ for the Green function, reducing
the problem to a system of ordinary differential equations ${\dot g}/g
= (1/2) \sum_k a_{kk}$ and ${\dot a}_{ij} = \sum_k a_{ik} a_{jk} + 2$.  We
assume the matrix $A \equiv \{a_{ij}\}$ to be replica symmetric, i.e.,
$a_{ii}=a$ and $a_{i\ne j}=b$ (this assumption implies no restriction,
as the present problem has a unique solution). The three resulting
differential equations ${\dot g}/g=na/2$, ${\dot b}=(n-2)b^2+2ab+2$,
and ${\dot a} = (n-1) b^2 + a^2 + 2$ are easily solved: concentrating
first on the difference $a-b$ we find, using the initial condition at
$t = 0$, the result $a-b = -1/t$. Using the Ansatz $c(t) =
\exp[-n\int^t dt' \, b(t')]$ in the equation for $b$, we can reduce the
problem to the Bessel equation and find the solution $b(t) = 1/nt - 2t
\cot(\sqrt{2n}t)/\sqrt{2n}t$, where we have again made use of the
initial condition. Finally, the result for $g(t)$ reads
$g(t)=(2n)^{1/4} \sqrt{t} \exp[-(1/4)\ln\sin^2(\sqrt{2n}t)]$ and
assembling the various elements, we arrive at the following final
expression for the moments ${\cal P}(n) = \langle Z^n(x,t)/Z_{\rm
th}^n(x,t) \rangle$ of the replicated partition function (the
arguments in ${\cal P}$ are $x\leftrightarrow u$ and $t\leftrightarrow
L$; we subtract the free energy $F_{\rm th}$ of the thermal model
($U=0$) as we are only interested in the difference $\Delta F = F -
F_{\rm th}$ with and without disorder; in the following we drop the
symbol `$\Delta$'),
\begin{equation}
{\cal P}(n) = \exp \! \bigg[\frac{n x^2}{2t}\! - \!\frac{1}{4}
\ln\frac{\sin^{2}(\sqrt{2n}t)} {2nt^2}\! -\! n^2 x^2 t
\frac{\cot(\sqrt{2n}t)}{\sqrt{2n}t}\bigg].
\label{Mellin_Transform}
\end{equation}

The moments (\ref{Mellin_Transform}) give us access to all
characteristic quantities of the random directed polymer on short
scales. In particular, the $n \rightarrow 0$ limit of
(\ref{Mellin_Transform}) determines the free energy average $\langle
F\rangle$ \cite{Parisi,Forgacs} and the probability function ${\cal
P}(u,L)$ for the displacement field $u$ \cite{Parisi}.  The large $n$
limit determines the `tails' ($F \to \pm \infty$) of the distribution
function ${\cal P}(F)$; the `left tails' ${\cal P}(F \to -\infty)$ for
the random directed polymer problem have been studied on short
\cite{Maritan,Halpin1} and long scales
\cite{Kardar,Zhang,Kolomeisky}. However, in order to determine the
`body' of the function ${\cal P}(F)$ one has to calculate {\it all}
the moments (\ref{Mellin_Transform}) (in the boson language the
determination of the `tails' and the `body' of the distribution
function ${\cal P}(F)$ amount to calculating the ground state of the
$n$-boson problem and the full spectrum, respectively). In the
following, we present a brief derivation of these quantities as they
follow from (\ref{Mellin_Transform}).

The $n \rightarrow 0$ limit of (\ref{Mellin_Transform}) determines the
disorder averaged free energy\cite{com} $\langle F\rangle=-T\langle\ln
[Z/Z_0]\rangle = -T \lim_{n \rightarrow 0}$ $[(Z/Z_0)^n-1]/n = -\alpha
L^2/12\epsilon$.  Similarly, this limit determines the correlation
functions for the displacement field $u$: These follow from the
probability function ${\cal P}(u,L) = \langle Z(u,L)/
\int_{-\infty}^{+\infty} du \, Z(u,L)\rangle =
\sqrt{\gamma/\pi}\exp(-\gamma u^2)$, where $\gamma^{-1} =
2(TL/\epsilon + \alpha L^3/3\epsilon^2)$.  The mean squared
displacement $\langle u^2(L) \rangle = \langle [u(L) - u(0)]^2
\rangle$ on scale $L$ then is given by $\langle u^2(L)
\rangle=(T/\epsilon) L + (\alpha/3\epsilon^2)L^3$ and crosses over
from diffusive thermal- to random wandering at a length scale $L \sim
\sqrt{\epsilon T/\alpha}$.

We turn to the calculation of the probability distribution function
${\cal P}(F)$. The moments ${\cal P}(n)$, see
(\ref{Mellin_Transform}), are nothing but the Mellin transform of
${\cal P}(F)$ \cite{Mellin},
\begin{equation}
{\cal P}(n) = \left \langle \frac{Z^n}{Z^n_{\rm th}} \right \rangle =
\int_{-\infty}^{+\infty} dF \, {\cal P}(F) \exp(-nF/T),
\label{pn}
\end{equation}
and we have reduced the problem of calculating ${\cal P}(F)$ to that
of inverting the Mellin transform. The latter depends only on the
combinations $\lambda \equiv 2t^2\, n = (\alpha L^2/\epsilon T)\, n$
and $\beta \equiv x^2/2t^3 = \epsilon^2 u^2/\alpha L^3$, i.e., we can
rewrite (\ref{pn}) in the form ($F_0 \equiv 2 T t^2 = \alpha L^2
/\epsilon$)
\begin{eqnarray}
f(\beta,\lambda(n))&=&\!\int_{-\infty}^{+\infty}\!\!\! dF\,{\cal P}(F)
\exp(-\lambda F/F_0) \quad \!\! (={\cal P}(n)) \label{plambda} \\ &=&
\exp{\bigg[\frac{\beta}{2}\lambda - \frac{1}{4}
\ln\frac{\sin^2(\sqrt{\lambda})}{\lambda} - \frac{\beta}{2}\lambda^2
\frac{\cot{(\sqrt{\lambda})}} {\sqrt{\lambda}}\bigg]}.\nonumber
\end{eqnarray}
The function $f(\beta, \lambda)$ is analytic in the halfplane ${\rm
Re}\{\lambda\} < \pi^2$ with $f(\beta,\lambda \rightarrow
{{\pi}^{2}}-0)\rightarrow\infty$.  We then can invert the Mellin
transform and write the function ${\cal P}(F)$ as an integral of $f$
over the imaginary axis, once we have found the proper analytical
continuation of $f(\beta, \lambda)$ (see below),
\begin{equation}
{\cal P}(F)= \frac{1}{2\pi F_0} \int_{-\infty}^{+\infty} d\lambda \,
f(\beta ,i\lambda) \exp(i\lambda F/F_0).
\label{lambdafunction}
\end{equation}
Note that the condition for the existence of the inverse Mellin
transform, $\int d\lambda \, | f(\beta,i\lambda)| < \infty$ is
satisfied: for $\beta=0$ we have $|f(0,i\lambda \rightarrow\pm
i\infty)| \sim \exp(-\sqrt{|\lambda|/2})$ while for $\beta \ne 0$,
$|f(\beta,i\lambda\rightarrow\pm i\infty)| \sim \exp(-\beta
|\lambda/2|^{3/2})$.

{\it Tails}: With $f(\beta, \lambda)$ analytic for ${\rm
Re}\{\lambda\} < \pi^2$ the exponential $\exp(-\lambda F/F_0)$, $F<0$,
in (\ref{plambda}) is compensated by the decaying tail of ${\cal
P}(F)$ up to the point $\lambda =\pi^2$, implying a left tail of the
form
\begin{equation}
{\cal P}(F\rightarrow -\infty) \sim \exp(-\pi^2 |F|/F_0).
\label{l_tail}
\end{equation}
The right tail of ${\cal P}(F)$ is determined by $f(\beta ,\lambda
\rightarrow -\infty)$.  Since $f(\beta \ne 0, \lambda\rightarrow
-\infty) \sim\exp(\beta|\lambda|^{3/2}/2)$ remains finite at finite
$|\lambda|$, we conclude that ${\cal P}(F)$ decays faster than 
exponential. We can invert the Mellin transform asymptotically via 
the method of steepest descents and obtain
\begin{equation}
{\cal P}(F \rightarrow \infty) \sim \exp(-16F^3/27\beta^2 F_0^3) 
\quad (\beta \ne 0).
\label{r_tail}
\end{equation}
The case $\beta =0$ has to be treated separately: With $f(\beta=0,
\lambda\rightarrow -\infty) \sim \exp(-\sqrt{|\lambda|}/2)$, we find
${\cal P}(F > 0)\equiv 0$. Using the Ansatz ${\cal P}(F) \propto
\exp(-A/|F|^\alpha)$ in combination with the method of steepest
descents produces the result
\begin{equation}
{\cal P}(F \rightarrow -0) \sim \exp(-F_0/16|F|)
\quad (\beta =0).
\label{r_tail_0}
\end{equation}

{\it Body}: We have to find the proper analytic continuation of the
Mellin transform (\ref{Mellin_Transform}) to imaginary $\lambda$
values (imaginary boson number), see (\ref{lambdafunction}).
Inspiration how to carry out this analytic continuation can be
obtained via the alternative route defining the distribution function
in terms of a path integral over the stochastic field $f$,
\begin{equation}
{\cal P}(F) = \int {\cal D}[f(z)] \, \delta (F-{\cal H}[u_0(f)]
+\epsilon u^2/2L),
\label{pathintegral}
\end{equation}
where $u_0(z)$ is the solution of the equation $\epsilon
u_0^{\prime\prime}(z) =-f(z)$ with the appropriate boundary conditions
$u_0(0)=0$ and $u_0(L)=u$ and ${\cal H}[u_{0}(z)]$ is the associated
energy ($(x,y)$ denotes the usual scalar product),
\begin{eqnarray}
&& \qquad\qquad {\cal H}[u_0(z)] = \frac{\epsilon
u^2}{2L}+\frac{1}{2\epsilon} (f,{\hat {\cal A}}f) - \frac{u}{L} (z,f),
\label{energyfu} \\
&& \,\,{\hat {\cal A}}f = \int_0^L dz^{\prime}\, \left[\frac{L-z}{L}
(L-z^\prime) + {\rm max}(z,z^\prime)-L\right] f(z^{\prime}). \nonumber
\end{eqnarray}
We determine the spectrum of the Hermitian operator $\hat {\cal A}$
($A_n = -L^2/\pi^2n^2$) and after integration over the stochastic
field $f$, we arrive at the distribution function in the form ${\cal
P}(F) = \int_{-\infty}^{+\infty} (d\mu/2\pi) \exp[i\mu F-g(\mu)]$,
with
\[
g(\mu) = \! \sum\limits_{n=1}^{\infty} \! \left[ \ln\left[1 \! - \!
\frac{i\alpha L^2\mu} {\epsilon\pi^2 n^2}\right]^{\frac{1}{2}} \!\!
+\frac{\alpha L u^2 \mu^2}{\pi^2} \left[n^2 \! - \! \frac{i\alpha
L^{2}\mu} {\epsilon\pi^2}\right]^{-1}\right]
\]
(defining $\mu = i\lambda/F_0$, a few algebraic manipulations
transform this result back to (\ref{plambda})). The above result
allows for a straightforward inversion of the Mellin transform via
numerical integration and thus to reconstruct the full distribution
function ${\cal P}(F)$, see Fig.\ 1. Furthermore, it implies that
${\cal P}(F>0)\equiv 0$ for $u =0$: We integrate over the contour in
the $\mu$-plane made from the real axis and the upper ($F>0$)
semi-circle. The integral over the arc vanishes asymptotically. With
all branching points situated in the lower half-plane, the contour
integral vanishes and ${\cal P}(F>0) \equiv 0$. This property of
${\cal P}(F)$ can be understood in terms of a minimization of the
functional ${\cal H}[u(z)]$: With the endpoint $(0,L)$, the test
function $u_t(z)\equiv 0$ satisfies the boundary conditions and we
have ${\cal H}[u_0] \leq {\cal H}[u_t] = 0$, hence ${\cal
P}(F>0)=0$. This analysis cannot be applied to other endpoints with $u
\ne 0$.

The distribution function ${\cal P}(F)$ for the Larkin model,
particularly its `tails', have been studied before and a 
comparative discussion is instructive. Expanding the correlator 
$K(u)$ in a Taylor series, $\langle U(u,z) U(u^\prime,z^\prime) 
\rangle \approx [K(0)-(\alpha/2)(u-u^\prime)^2]\delta (z-z^\prime)$,
the authors of Refs. \onlinecite{Maritan,Halpin1} arrive at
the asymptotic expression $\langle Z^n\rangle \sim 
\exp[- C\sqrt{n}(n-1)L]$, which cannot be identified with the
Mellin transform of some distribution function ${\cal P}(F)$,
however. Keeping only the second term $\exp(C\sqrt{n}L)$,
as suggested in Ref.\ \cite{Halpin1}, is questionable: though the 
left tail $\sim \exp(-BL^2/|F|)$ reproduces the desired answers 
for the fluctuations of the free energy $\delta F\propto L^2$
and the line wandering $u\propto L^{3/2}$, this tail is 
unacceptable for a properly normalizable distribution function. 
The origin of the problem seems to lie in the correlator expansion 
itself: determining the exact Mellin transform using the expanded 
correlator we find the result 
\[
{\cal P}_{\rm exp} (n) = 
\exp\bigg[\frac{n-1}{4}\ln\frac{2nt^{2}}{\sinh^{2}
\left(\sqrt{2n}t\right)}
+\frac{{\tilde K}}{2}n^{2}\bigg],
\]
where ${\tilde K}=K(0) L/T^2$. While the leading term has the
correct sign, the fact that ${\cal P}_{\rm exp} (n = -\pi^2/2t) = 0$
implies that ${\cal P}_{\rm exp}(F)$ assumes negative values, in
conflict with the required positivity of ${\cal P}$. 
The findings of Parisi \cite{Parisi} show that the expansion 
provides correct results for the mean value of the free energy 
and the displacement correlation function, however, higher moments 
emphasizing larger distances $u-u^\prime$, where the expanded 
correlator becomes negative, turn out wrong and so does the 
function ${\cal P}(F)$. Carrying the expansion of the potential 
further, $U(u,z) \approx U(0,z) - f(z)u + g(z)u^2/2$,
the calculation of the distribution function ${\cal P} (F)$ 
can be done along the lines described above (see 
(\ref{pathintegral})) and we recover the result for the Larkin 
model for $\beta = 0$, see Fig.\ 1: as expected, the answer for 
the translation invariant random polymer problem (\ref{Hamiltonian}) 
and (\ref{correlr}) with a homogeneous correlator $K(|u-u'|)$ is 
independent of the coordinate $u$ of the polymer's endpoint.

Next, we address the {\it large scale} behavior of the random directed
polymer, which can be solved via the Bethe Ansatz technique
\cite{Kardar}. The ground state energy $E_0(n)\propto -n^2(n-1)$
produces the left tail $\ln{\cal P}(F)\propto -|F|^{3/2}/L^{1/2}$ and
provides the scaling $\delta F\propto L^{{1}/{3}}$ for the
fluctuations in $F$ and the correct exponent $\zeta =2/3$. Still, the
generalization of this procedure seems not straightforward.  E.g., for
the $(1+2)$-problem the ground state energy increases faster than any
power of $n$, implying that $\delta F$ behaves logarithmically and
$\zeta =1/2$, in contradiction with the expected superdiffusive
($\zeta >1/2$) behavior of the $(1+2)$-dimensional string. It then
seems important to know the complete Mellin transform ${\cal P}(n)$
for an accurate calculation of the wandering exponent $\zeta$.

An interesting remark concerns the wandering exponent $\zeta = 3/2$ 
for the Larkin problem and its relation with the Kolmogorov scaling 
for the small-distance velocity fluctuations in the Burgers 
turbulence problem, $\langle [v(u,L=\infty)- v(-u,L=\infty)]^3\rangle 
\propto u^{3\nu}$, with $\nu = 1/3$ (see also \cite{Mezard}). The
consistency between the two exponents $\zeta$ and $\nu$ is easily
checked: The fields in the two problems are related via $v =
\partial_u F \sim F/u$, hence $F \propto u^{1+\nu}$. In the Larkin
model $F \propto L^2 \propto u^{2/\zeta}$ and we obtain $2/\zeta =
1+\nu$, which is indeed satisfied.  However, this result does not
carry a deep physical meaning but is simply a consequence of
scaling. Physically, the two problems are very different: In the
Larkin model we study the free energy $F$ both as a function of $u$
and (finite) $L$ with only one pinning valley relevant, whereas for
the case of Burgers turbulence the stationary correlation functions
$m_{k}(u) = \langle [v(u, L=\infty)-v(-u,L=\infty)]^k\rangle$ are
studied, where many disorder-induced valleys are relevant (in the
language of the directed polymer problem).

In the end, the calculation of the probability distribution ${\cal
P}[F(u,L)]$ turns out to be quite a non-trivial problem: a successful
attack on the problem requires the {\it precise} knowledge of {\it
all} the moments ${\cal P}(n)$. Under these circumstances, we then can
proceed with the analytical continuation to imaginary boson number and
invert the Mellin transform to obtain ${\cal P}(F)$.  The successful
application of this strategy to the Larkin model uncovers the
limitations of previous attacks on the problem and may serve as a
guideline for future studies.

We thank Thomas Nattermann and Valerii Vinokur for helpful
discussions.

\end{multicols}          %Comment out for Latex209


\begin{thebibliography}{99}
\bibitem{DR} We omit the term $\int_0^L U(0,z) \thinspace dz$
corresponding to the trivial part of disorder.  The Mellin transform
of the probability distribution for the full free energy is given by
the product ${\cal P}(n) \exp (K(0)Ln^2/2T^2)$.
\bibitem{Mellin} The Mellin transform $f(p)$ of the function $f(x)$ is
defined as $f(p)=\int_0^\infty dx \, f(x)x^{p-1}$.  With $x=\exp(-t)$
this transforms to $f(p)=\int_{-\infty}^{+\infty} dt \,
f(\exp(-t))\exp(-pt)$.
\end{thebibliography}
\end{document}